\documentclass{jpp}
\usepackage{graphicx}
\usepackage{amsmath}
\usepackage{epstopdf, epsfig}

\shorttitle{3D MHD Equilibria}
\shortauthor{S. A. Lazerson}

\title{From tokamaks to stellarators: understanding the role of 3D shaping\footnote{Notice: This manuscript has been authored by Princeton University under Contract Number 
DE-AC02-09CH11466 with the U.S. Department of Energy. The publisher, by accepting the article for publication acknowledges, that the United States Government retains a non-exclusive, paid-up, irrevocable, world-wide license to publish or reproduce the published form of this manuscript, or allow others to do so, for United States Government purposes.}}

\author{Samuel A. Lazerson\aff{1}
  \corresp{\email{slazerso@pppl.gov}} \and John C. Schmitt\aff{2}}

\affiliation{\aff{1}Princeton Plasma Physics Laboratory,
Princeton, NJ 08543, USA
\aff{2}Auburn University,
Auburn, AL 36849, USA}

\begin{document}

\maketitle

\begin{abstract}
In this work, the role which three-dimensional shaping plays in the generation of rotational transform in toroidal magnetically confinement devices is explored.  The susceptance matrix as defined by \cite{Strand:2001cd} is presented and compared to simulations of three dimensional MHD equilibria.  The dependence of the edge rotational transform on axisymmetric shaping is briefly explored.  It is then shown how simple three dimensional shaping can drive rotational transform in a simply shaped stellarator.  It is found that modes with high poloidal mode number drive edge transform with little effect on the core, while modes with poloidal mode $m<2$ drive both edge and core rotational transforms. This work concludes with a discussion of the non-linearity of the three dimensional shape space as defined in terms of magnetic symmetry.
\end{abstract}

\section{Introduction}
The goal of magnetic confinement nuclear fusion is to use magnetic fields to confine a hot plasma long enough so that a burning plasma state can be reached.  While many magnetic configurations have been considered, the topological torus shows the greatest maturity.  In this class of device two subclasses exist, those which are axisymmetric and those which are non-axisymmetric.  In the axisymmetric branch, exist the tokamak and field reversed configuration.  While in the non-axisymmetric branch one finds stellarators and heliotrons.  Spanning the space are the reverse field pinch devices which can exhibit an axisymmetric nature, or bifurcate into a state with non-axisymmetric plasma cores.  In this paper, the role of three-dimensional shaping in confinement is explored.

\section{Theory}
 Consider a  charged particle in a uniform magnetic field.  The relevant force acting on the particle is the Lorentz force
\begin{equation}
  \vec{F}=q\left(\vec{E}+\vec{v}\times\vec{B}\right)
  \label{Lorentz}
\end{equation}
 where $q$ is the particles charge, $\vec{E}$ the electric field, $\vec{v}$ its velocity, and $\vec{B}$ is the magnetic field.  If the particle possesses any component of velocity perpendicular to the magnetic field it will begin to exhibit cyclotron motion.  One may generate such a field using a solenoid, however collisions between particles will tend to equalize the components of the velocity parallel and perpendicular to the magnetic field.  Thus particles will tend to stream along the magnetic field line at finite temperatures (ignoring an additional hierarchy of complicated physics).  This implies that we must in some way `cap' the ends of this magnetic bottle.
 
 One concept for `capping' the bottle is to use a magnetic mirror force to reflect the particles back along the field line.  However, this is only a partial solution as the fastest particle will still tend to escape.  Ultimately cooling the plasma before being able to reach fusion relevant temperatures.  The more agreed upon solution is to deform the straight field into a topological torus.  While this avoids the problems of end losses, the magnetic field now becomes curved (and a field gradient forms).  This curvature of the magnetic field causes particles to drift across magnetic field lines.  For particles on the inboard side of the torus, such a drift brings them toward the center of the device. While for particles on the outboard side, the motion is away from the center.
 
 If the magnetic field lines could be curved in such a way that particles spend more of their time in the `good' curvature region than in the `bad,' the net effect of the magnetic field would be to `confine' particles to the hot core region.  The `Hairy Ball Theorem' of algebraic topology \citep{Eisenberg:1979hm} tells us that for a torus this is possible.  From this point forward we will restrict ourselves to a discussion where nested surfaces of toroidal flux exist everywhere in our magnetic configuration up to some enclosing flux surface.  This enclosing surface will be called the last closed magnetic surface (LCMS).  At the center of the torus we thus have a single magnetic field line.  While this assumption may in fact be artificial, the ramifications of such an assumption will be discussed later in this work.
 
 The `rate' at which magnetic field lines rotate about the magnetic axis is defined by a quantity called the rotational transform, which is defined in terms of poloidal ($\Psi$) and toroidal ($\Phi$) magnetic fluxes \citep{Bateman:1973kb,Strand:2001cd}
\begin{equation}
  \iota = \frac{d\Psi}{d\Phi}.
  \label{Iota}
\end{equation}
So, put another way, the act of confining charged particles with magnetic fields in a toroidal device is the act of generating rotational transform.

\begin{figure}
  \centerline{\includegraphics[width=\textwidth]{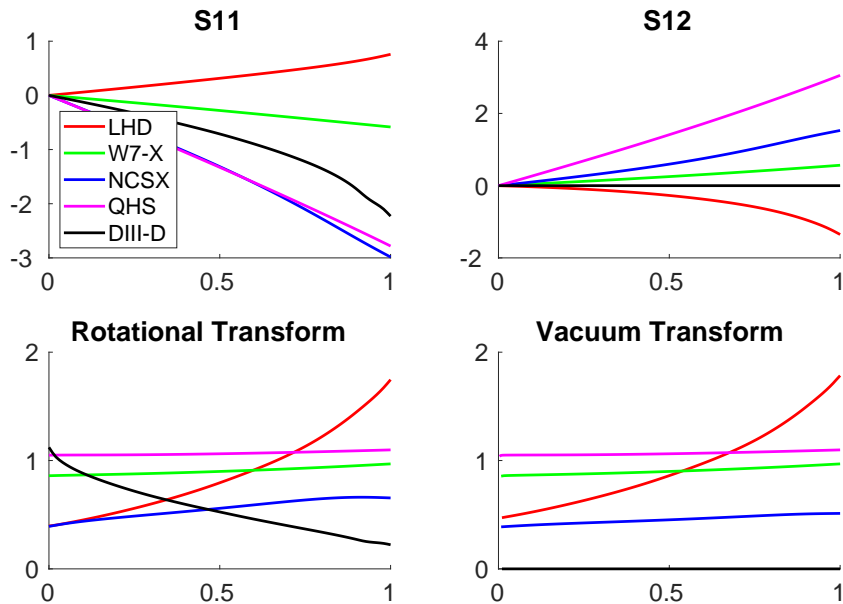}}
  \caption{Susceptance coefficients for various magnetic confinement devices.  The $x$-axis is in terms of normalized toroidal flux.}
\label{fig:S_comp}
\end{figure}

Rotational transform can be generated by non-axisymmetric shaping or toroidal current as defined by
\begin{equation}
  \iota = \frac{\mu_0 I }{S_{11} \Phi'}-\frac{S_{12}}{S_{11}}
  \label{Iota_gen}
\end{equation}
where $\mu_0$ is the permeability of free space, $I$ the toroidal current, $\Phi'=d\Phi/d\rho$, and $S$ are the components of the susceptance matrix.  In general the suseptance matrix is defined as:
\begin{equation}
  \mu_0 \begin{pmatrix} I  \\ F \end{pmatrix} = \begin{pmatrix} S_{11} & S_{12} \\ S_{21} & S_{22}\end{pmatrix}\begin{pmatrix} d\Psi/d\rho \\ d\Phi/d\rho \end{pmatrix}
  \label{S_eq}
\end{equation}
where $F$ is the poloidal current and $\rho$ is a radial coordinate.  Assuming a nested flux surface representation and a magnetic stream function $\lambda$ we can redefine our poloidal angle ($\theta^*=\theta+\lambda\left(\rho,\theta,\zeta\right)$), then allowing one to write the magnetic field in the form
\begin{equation}
  \vec{B}=\frac{1}{2\pi\sqrt{g}}\left[\left(\frac{d\Psi}{d\rho}-\frac{d\lambda}{d\zeta}\frac{d\Phi}{d\rho}\right)\vec{e}_\theta+\left(1+\frac{d\lambda}{d\theta}\right)\frac{d\Phi}{d\rho}\vec{e}_\zeta\right].
  \label{B_eq}
\end{equation}
where $\sqrt{g}$ is the coordinate Jacobian.  Here we note that our choice of magnetic field representation is exactly that of the VMEC equilibrium code \citep{1983PhFl...26.3553H}, allowing us to write the components of the susceptance matrix in terms of geometric quantities
\begin{equation}
  S_{11}=\frac{dV/d\rho}{4\pi^2}\left<\frac{g_{\theta\theta}}{g}\right>
  \label{S_11}
\end{equation}
\begin{equation}
  S_{12}=\frac{dV/d\rho}{4\pi^2}\left<\frac{g_{\theta\zeta}\left(1+d\lambda/d\theta\right)-g_{\theta\theta}d\lambda/d\zeta}{g}\right>
  \label{S_12}
\end{equation}
\begin{equation}
  S_{21}=\frac{dV/d\rho}{4\pi^2}\left<\frac{g_{\zeta\zeta}\left(1+d\lambda/d\theta\right)-g_{\zeta\theta}d\lambda/d\zeta}{g}\right>
  \label{S_21}
\end{equation}
\begin{equation}
  S_{22}=\frac{dV/d\rho}{4\pi^2}\left<\frac{g_{\zeta\theta}}{g}\right>
  \label{S_22}
\end{equation}
where $g_{kk}$ are the components to the coordinate metric tensor and $V$ is the volume enclosed in a flux surface.  Clearly the $S_{11}$ and $S_{22}$ coefficients depend purely on the geometry of the flux surfaces.  Figure \ref{fig:S_comp} depicts the values of the $S_{11}$ and $S_{12}$ coefficients for a few common MHD equilibria.  Note that at the edge ($x=1$) the W7-X device has a rather small value of $S_{11}$.  This suggests a strong sensitivity of the edge rotational transform to enclosed toroidal current (an already well documented issue) \citep{BOZHENKOV20132997}.  The relatively large values of $S_{11}$ in the NCSX and QHS configurations suggest a robustness of the configurations to current (an already documented feature) \citep{doi:10.1063/1.4978494}.

\section{The dependence of rotational transform on flux surface geometry}
Equation \ref{Iota_gen} tells us that there are two ways of generating transform: toroidal current and three-dimensional shaping.  To explore this statement we employ the VMEC equilibrium code to calculate various equilibria and assess the impact of magnetic geometry on rotational transform.  To do this we first examine the effect of the $g_{\theta\theta}$ term in axisymmetry then move onto non-axisymmetric perturbations.

As VMEC will be our tool it is worthwhile to review the nature of the flux surface representation used in the code.  For stellarator symmetric equilibrium the flux surfaces are parametrized by functions which trace out the magnetic surfaces, namely
\begin{equation}
  R\left(\rho,\theta^*,\zeta\right) = \sum_{n=-N}^N\sum_{m=0}^M R\left(\rho\right)\cos{\left(m\theta^*-nN_{FP}\zeta\right)}
  \label{R_eq}
\end{equation}
\begin{equation}
  Z\left(\rho,\theta^*,\zeta\right) = \sum_{n=-N}^N\sum_{m=0}^M Z\left(\rho\right)\sin{\left(m\theta^*-nN_{FP}\zeta\right)}
  \label{Z_eq}
\end{equation}
where $N$ is the toroidal truncation of modes, $M$ the poloidal truncation of modes, and $N_{FP}$ the magnetic field periodicity.  The $\lambda\left(\rho,\theta^*,\zeta\right)$ function has odd parity, thus a similar form to equation \ref{Z_eq}.

\subsection{Axisymmetric equilibria}
\begin{figure}
  \centerline{\includegraphics[width=\textwidth]{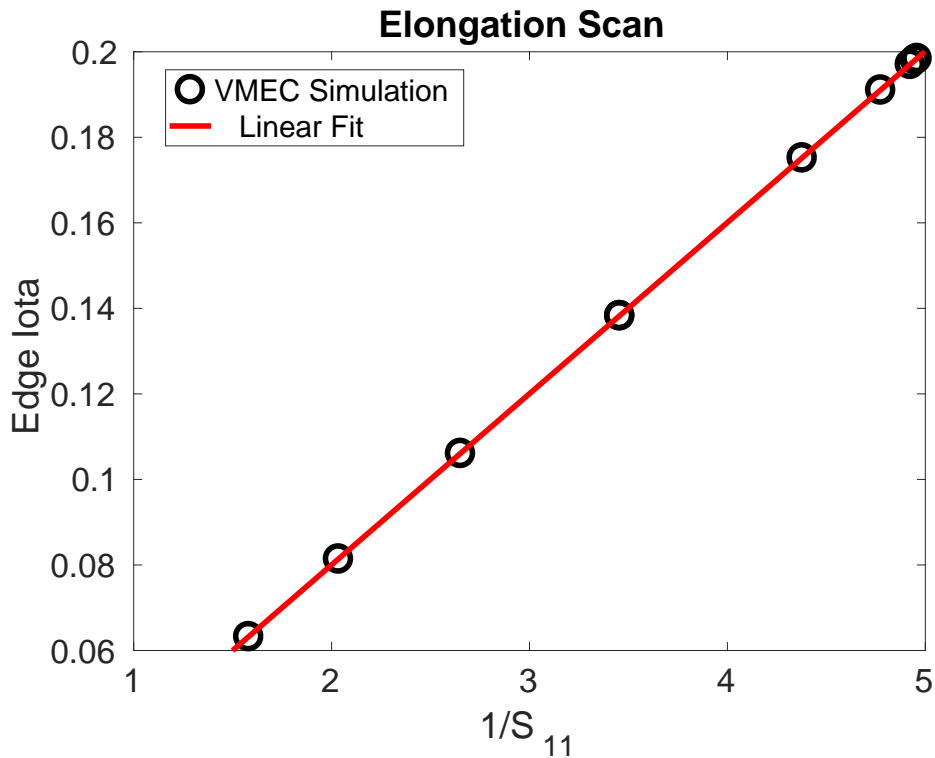}}
  \caption{Comparison between edge rotational transform and $S_{11}$ component of the susceptance matrix showing the scaling relationship.  Circles indicate simulation results while the line is a linear fit to said data. Results become more elliptical going right to left.}
\label{fig:axi_comp}
\end{figure}

For axisymmetric equilibria the $S_{12}$ component of the susceptance matrix vanishes, leaving only the $S_{11}$ component (and toroidal current) to generate rotational transform.  An aspect ratio $10$ circular cross section equilibrium (at zero beta but finite toroidal current) serves as a basis for examining the relationship between $S_{11}$ and rotational transform.  It is then a simple matter to elliptically deform the equilibria by adjusting the $n=0,m=1$ mode of the $Z$ harmonics.  Figure \ref{fig:axi_comp} depicts the dependence of edge transform on the $S_{11}$ component of the susceptance matrix as the equilibrium is deformed.  In this plot the equilibria with larger deformation are located on the left of the graph.  It is important to remember that rotational transform is the inverse of the safety factor ($\iota=1/q$).  The result tells us that as we add ellipticity to the axisymmetric equilibrium the value of $q$ increases (holding all else constant).

If we consider the effect on the iota profile, this result holds for other modes which increase the $g_{\theta\theta}$ metric.  The toroidal current to zeroth order acts as a rigid shift to the rotational transform (ignoring profile effects).  While the boundary perturbations have more nuanced effects on the rotational transform profile.  The $m=1$ modes (which generate ellipticity) tend to penetrate to the core, meanwhile the higher order modes tend to decay into the core with higher orders falling off faster.  This makes it possible to tailor the iota profile through shaping and current.  This implies that multiple combinations of current profile and axisymmetric shaping may be employed to generate a given transform profile.  

\subsection{Non-axisymmetric equilibria}
When the toroidal symmetry of the device is broken the $S_{12}$ term becomes non-zero allowing rotational transform to be produced even when the enclosed toroidal current becomes zero.  Such is the design concept surrounding stellarators and heliotrons.  Here the ratio $\iota_{vac}=-S_{12}/S_{11}$ defines the `vacuum' transform.  It should now be clear that by maximizing the $S_{11}$ component of the susceptance, a robustness of the iota profile to currents (be they ohmic, bootstrap, beam driven or electron-cyclotron drive) can be built into an equilibrium.  However, this also suggests that the $S_{12}$ component be maximized as well to generate the vacuum transform.

\begin{figure}
  \centerline{\includegraphics[width=\textwidth]{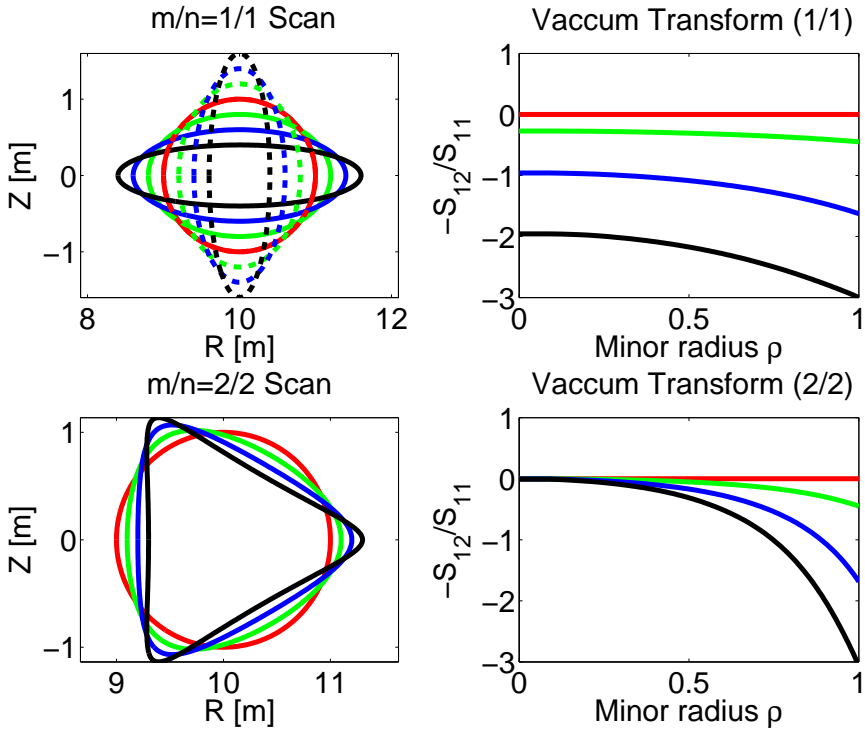}}
  \caption{Examples of non-axisymmetric modes driving vacuum rotational transform.  Colors relate boundary shapes to transform profiles while dashed line indicate boundary shape at the second symmetry plane.  The negative values are due to choice of sign on poloidal angle and toroidal harmonics.}
\label{fig:nonaxi_comp}
\end{figure}

A better understanding of how boundary shapes drive vacuum transform in stellarators can be seen in figure \ref{fig:nonaxi_comp}.  We begin by perturbing the axisymmetric equilibria (toroidal current set to zero) with $n=1,m=1$ (upper plots) and $n=2,m=2$ perturbations (lower plots) and examine the resulting rotational transform profiles.  The $n=1,m=1$ perturbation deforms the equilibria into a rotating ellipse.  The resulting rotational transform profile posses reversed shear and modifies both the edge and axis values of the profile.  The $n=2,m=2$ perturbation creates a rotating triangular feature to the equilibrium, becoming slightly scalloped at the largest amplitude explored.  It is worthwhile to note that the $n=2$ nature implies that the half-field period plane has the same cross section as the other symmetry planes.  The $m=2$ nature of this perturbation results in a modification to the rotational transform at the edge but not in the core.  We note that the sign of these transform profiles is a function of the choice of sign in the toroidal modes.

While these two examples appear trivial, they already show that three dimensional shaping can be used to tailor the vacuum rotational transform of a stellarator.  Specifically the low order modes ($m<2$) should provide an ability to tailor the transform on axis.  Meanwhile the higher order modes allow tailoring of the edge transform.  Thus the design of a stellarator becomes a the task of combining these modes to produce an equilibrium with the desired properties.

\begin{figure}
  \centerline{\includegraphics[width=\textwidth]{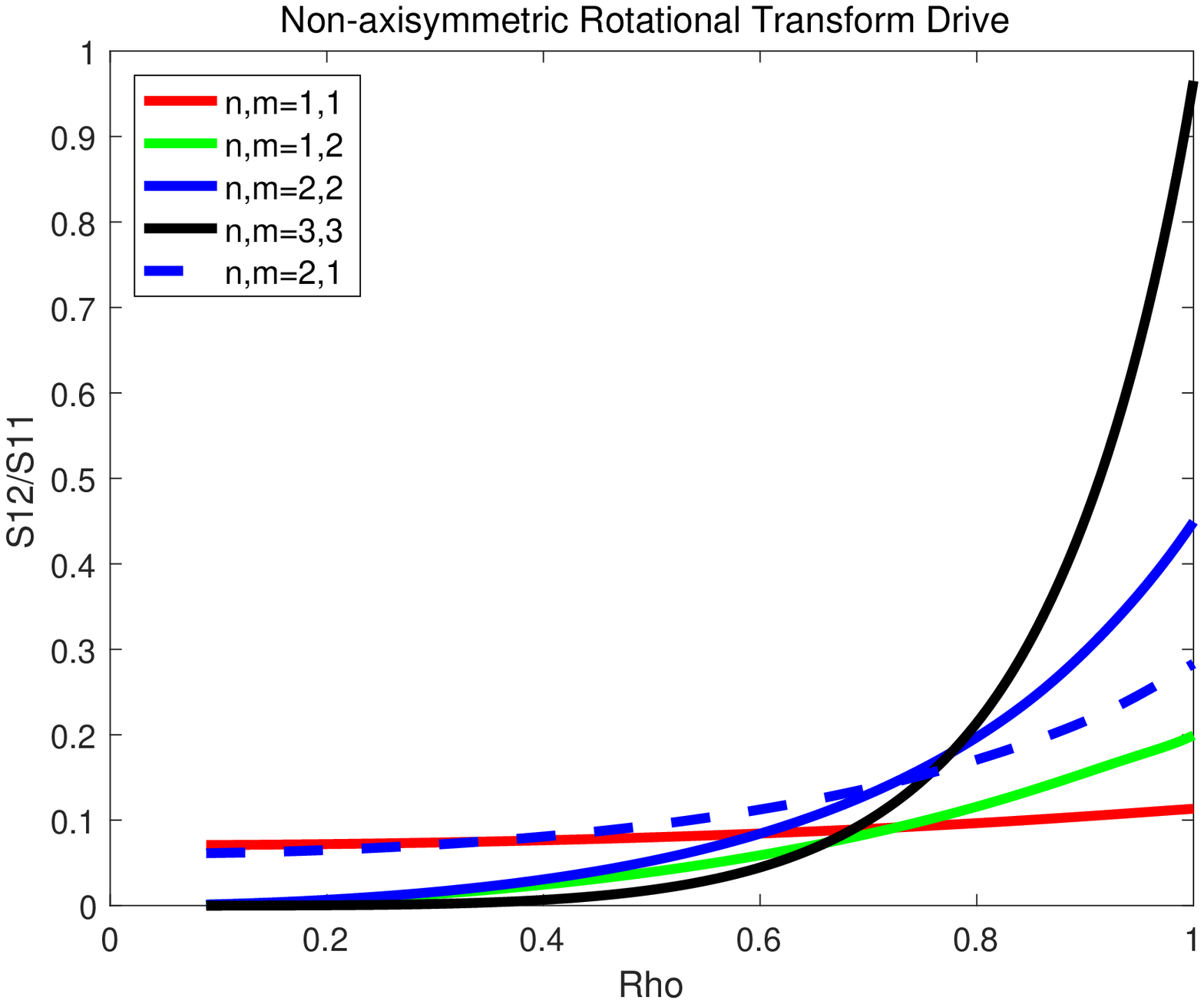}}
  \caption{Comparison of vacuum transform driven by various non-axisymmetric modes.  Care is taken so that edge mode amplitude is similar in each case.}
\label{fig:mod_drive}
\end{figure}

Stellarators usually use a multitude of modes to drive rotational transform.  In figure \ref{fig:mod_drive}, a set of boundary deformations is explored where the relative amplitudes of the modes are equal.  We see that modes with $m=1$ nature penetrate to the core region while modes of higher m are damped.  By changing the sign of the modes the sign of their contribution to the rotational transform can also be changed.  We see that the $n=1,m=2$ mode drive a similar level of core transform as that of the $n=1,m=1$ mode but significantly more edge transform.  Simulations indicate that it is possible to combine the $n=1,m=1$ boundary modes and $n=2,m=2$ boundary modes and recompute the equilibria.  The resulting equilibria has a transform almost identically matching that found though summation of the two individual transforms.  This result may not hold should the magnetic axis become sufficiently non-axisymmetric.

\section{Discussion}
This work has begun to explore the role of three dimensional shaping on the generation (and modification) of rotational transform in ideal magnetohydrodynamic toroidal equilibria.  While not exhaustive, this work shows that modes with poloidal modes less than $2$ are the primary means by which transform on axis can be driven.  The dependence of core transform on edge poloidal perturbations appears weak at best, with the $n=1,m=1$ mode driving nearly the same amount of transform as the $n=2,m=1$ mode for similar perturbation amplitudes.  At the edge of the equilibrium modes with higher poloidal mode number appear to drive larger amounts of transform for the same level of boundary displacement.  However, this suggests that the same (or similar) rotational transform profiles can be achieved through a multitude of boundary modes.  Numerical simulations combining different modes supports the idea of linear combinations of modes to produce a profile which is the the sum of the profiles driven by different modes.

This work suggests a new target for stellarator optimization.  Specifically if a stellarator was optimized in such a way to maximize the $S_{11}$ coefficient, the resulting configuration would become robust to current changes.  This would provide rotational transform robustness to bootstrap currents.  This is desirable in stellarators which employ island divertors (such as Wendelstein 7-X).  These devices rely on an edge island chain of known position and width.  Thus these devices would benefit from edge iota profile rigidity.  One metric could be defined by the transform profile rigidity as
\begin{equation}
R=\frac{\delta P}{\delta I} = \frac{\mu_0/\Phi' - \partial S_{12}/\partial I}{\mu_0 I/\Phi' -S_{12}}
\end{equation}
where we have defined low values of rigidity ($R$) to indicate a lack of sensitivity to toroidal currents.  The variable $P$ is used to represent the normalized change in rotational transform for a given unit current.  It should be noted that the partial derivative ($\partial S_{12} / \partial I$) can become quite complex and as such we leave leave further analysis to future work.

It is important to note that in this work, we have ignored the possibility of islands and stochastic regions at rational surfaces.  Moreover, we have completely neglected the recent results by \cite{Loizu:2015ig} regarding the necessity of non-smooth rotational transform profiles across rational surfaces.  If we imagine a plasma model which allows the islands to open, then we expect the island width to scale as the resonant harmonic.  While modifying the profile in the neighborhood of the island, overall the equilibrium profile is mostly unperturbed (excluding pathological examples).  We note that there clearly exists a freedom in approximating a rotational transform profile using many different modes.  This suggest that one could design a stellarator by first choosing a rotational transform profile with known resonances, then explicitly setting the resonant component to zero and using the non-resonant boundary deformations to create the transform.

If we consider inclusion of shielding currents, then our equilibrium rotational transform profile becomes discontinuous at each resonant rational surface.  These discontinuities scale as the amplitude of the resonant harmonic and should be small in most cases.  The resulting equilibria are only modified in the vicinity of these surfaces and should not impact the generic features of the equilibrium \citep{Lazerson:2016hv}.  One should note that such shielding currents require toroidal sheet currents and as such are inconsistent with the zero beta, zero current equilibrium limit.

\begin{figure}
  \centerline{\includegraphics[width=\textwidth]{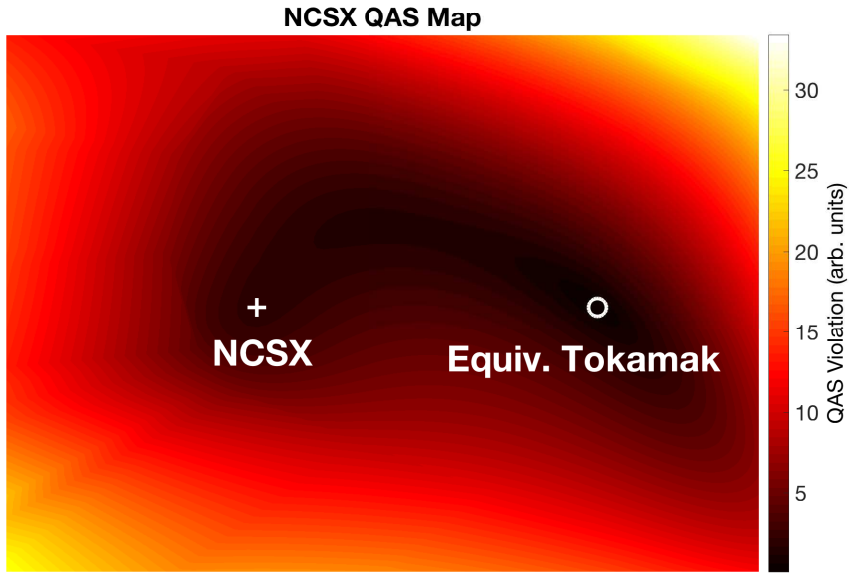}}
  \caption{Mapping on n-dimensional parameter space around the NCSX fixed boundary equilibium (+).  An axisymmetric tokamak equilibria (similar to NCSX, o) and a tubulent transport optimized version of NCSX (located just off plot vertically from NCSX).  Color contours indicate levels of non-axisymmetry with black indicating perfect symmetry and white indicating larger violations of magnetic quasi-axisymmetry.  The axes are eigenvectors in the shape space.  The $x$ axis going from axisymmetry to NCSX, while the $y$ axis goes from NCSX to it's transport optimized version.}
\label{fig:ncsx_map}
\end{figure}

Before concluding this discussion the authors would like to take the time to point out that generation of transform alone is not enough to confine the particles in a nuclear fusion device.  To this end a great deal of work (mostly associated with stellarators) has gone into the development of optimized three-dimensional equilibria \citep{Mynick:1982fg,1998PhPl....5.1752S, Mynick:2006bt}.  Here the optimization usually focuses on minimizing the radial transport of particles through magnetic symmetries.  These symmetries include helically symmetric (the HSX device) \citep{Boozer:1995fb}, quasi-oniginatiy (W7-X) \citep{Hall:1975ey}, and quasi-axisymmetry (NCSX) \citep{Garabedian:1996hh}.  NCSX being relevant to this paper as the tokamak falls into the quasi-axisymmetric (QAS) category (axisymmetry being the exact realization of quasi-axisymmetry).  Of course any perturbation to a tokamak which breaks toroidal symmetry takes the equilibrium away from it's ideal axisymmetric state.  However, the situation in practice is quite more complicated.  In figure \ref{fig:ncsx_map}, we have mapped the n-dimensional space around the NCSX equilibrium by defining two vectors and evaluating equilibria in the subspace defined by these two vectors.  In one direction (along x-axis) we smoothly deform the NCSX equilibria into an axisymmetric tokamak equilibria.  In the other direction, we smoothly deform the boundary toward a turbulent transport optimized version of NCSX \citep{Mynick_1}.  The amplitude of the color map indicates how non-axisymmetric the magnetic equilibria are,  a sum over $|B|$ modes with toroidal mode numbers other than zero (normalized to the amplitude of those modes with $n=0$.)  This database was generated from $\sim10000$ VMEC runs produced by the STELLOPT code.  

The purpose of showing this seemingly separate work is to demonstrate that in moving from an axisymmetric tokamak to the quasi-axisymmetric NCSX equilibrium the non-axisymmetry increases and then decreases.  And while NCSX is clearly not axisymmetric, it is predicted to have transport levels approaching that of a tokamak.  It should now be clear that the selection of modes to generate transform in stellarators is much more complicated than the analysis presented here.  Still this work acts as a starting point, and future works will explore the dependency of radial transport, stability, and bootstrap current in future works.

\section{Citations and references}

The authors would like to acknowledge fruitful discussions with J. Loizu, S. Hudson, A. Boozer, H. Mynick, N. Pomphrey, S. Prager, and D. Gates.  Furthermore, the authors would like to thank S. Hirshman, S. Seal, and M. Cianciosa for access to the VMEC and PARVMEC codes.

\bibliographystyle{jpp}

\bibliography{Lazerson_3dequil}

\begin{thebibliography}{15}
\expandafter\ifx\csname natexlab\endcsname\relax\def\natexlab#1{#1}\fi
\def\au#1{#1} \def\ed#1{#1} \def\yr#1{#1}\def\at#1{#1}\def\jt#1{\textit{#1}}
  \def\bt#1{#1}\def\bvol#1{\textbf{#1}} \def\vol#1{#1} \def\pg#1{#1}
  \def\publ#1{#1}\def\arxiv#1{#1}\def\org#1{#1}\def\st#1{\textit{#1}}

\bibitem[Bader {\em et~al.\/}(2017)Bader, Boozer, Hegna, Lazerson \&
  Schmitt]{doi:10.1063/1.4978494}
{\sc \au{Bader, A.}, \au{Boozer, A.~H.}, \au{Hegna, C.~C.}, \au{Lazerson,
  S.~A.} \& \au{Schmitt, J.~C.}} \yr{2017}  \at{Hsx as an example of a
  resilient non-resonant divertor}.  \jt{Physics of Plasmas}  \bvol{24}~(3),
  \pg{032506},  \arxiv{arXiv: http://dx.doi.org/10.1063/1.4978494}.

\bibitem[Bateman(1973)]{Bateman:1973kb}
{\sc \au{Bateman, G.}} \yr{1973}  \at{{Energy principle with specific
  inductance}}.  \jt{Nuclear Fusion}  \bvol{13}~(2),  \pg{227--238}.

\bibitem[Boozer(1995)]{Boozer:1995fb}
{\sc \au{Boozer, A.~H.}} \yr{1995}  \at{{Quasi-helical symmetry in
  stellarators}}.  \jt{Plasma Physics and Controlled Fusion}  \bvol{37}~(11A),
  \pg{A103--A117}.

\bibitem[Bozhenkov {\em et~al.\/}(2013)Bozhenkov, Geiger, Grahl, Kißlinger,
  Werner \& Wolf]{BOZHENKOV20132997}
{\sc \au{Bozhenkov, S.}, \au{Geiger, J.}, \au{Grahl, M.}, \au{Kißlinger, J.},
  \au{Werner, A.} \& \au{Wolf, R.}} \yr{2013}  \at{Service oriented
  architecture for scientific analysis at w7-x. an example of a field line
  tracer}.  \jt{Fusion Engineering and Design}  \bvol{88}~(11),  \pg{2997 --
  3006}.

\bibitem[Eisenberg \& Guy(1979)]{Eisenberg:1979hm}
{\sc \au{Eisenberg, M.} \& \au{Guy, R.}} \yr{1979}  \at{{A Proof of the Hairy
  Ball Theorem}}.  \jt{The American Mathematical Monthly}  \bvol{86}~(7),
  \pg{571}.

\bibitem[Garabedian(1996)]{Garabedian:1996hh}
{\sc \au{Garabedian, P.~R.}} \yr{1996}  \at{{Stellarators with the magnetic
  symmetry of a tokamak}}.  \jt{Physics of Plasmas}  \bvol{3}~(7),
  \pg{2483--2485}.

\bibitem[Hall \& McNamara(1975)]{Hall:1975ey}
{\sc \au{Hall, L.~S.} \& \au{McNamara, B.}} \yr{1975}  \at{{Three-dimensional
  equilibrium of the anisotropic, finite-pressure guiding-center plasma: Theory
  of the magnetic plasma}}.  \jt{Physics of Fluids}  \bvol{18}~(5),
  \pg{552--15}.

\bibitem[Hirshman \& Whitson(1983)]{1983PhFl...26.3553H}
{\sc \au{Hirshman, S.~P.} \& \au{Whitson, J.~C.}} \yr{1983}
  \at{{Steepest-descent moment method for three-dimensional magnetohydrodynamic
  equilibria.}}  \jt{Physics of Fluids}  \bvol{26}~(12),  \pg{3553--3568}.

\bibitem[Lazerson {\em et~al.\/}(2016)Lazerson, Loizu, Hirshman \&
  Hudson]{Lazerson:2016hv}
{\sc \au{Lazerson, S.}, \au{Loizu, J.}, \au{Hirshman, S.~P.} \& \au{Hudson,
  S.~R.}} \yr{2016}  \at{{Verification of the ideal magnetohydrodynamic
  response at rational surfaces in the VMEC code}}.  \jt{Physics of Plasmas}
  \bvol{23}~(1),  \pg{012507}.

\bibitem[Loizu {\em et~al.\/}(2015)Loizu, Hudson, Bhattacharjee, Lazerson \&
  Helander]{Loizu:2015ig}
{\sc \au{Loizu, J.}, \au{Hudson, S.~R.}, \au{Bhattacharjee, A.}, \au{Lazerson,
  S.} \& \au{Helander, P.}} \yr{2015}  \at{{Existence of three-dimensional
  ideal-magnetohydrodynamic equilibria with current sheets}}.  \jt{Physics of
  Plasmas}  \bvol{22}~(9),  \pg{090704}.

\bibitem[Mynick(2006)]{Mynick:2006bt}
{\sc \au{Mynick, H.~E.}} \yr{2006}  \at{{Transport optimization in
  stellarators}}.  \jt{Physics of Plasmas}  \bvol{13}~(5),  \pg{058102}.

\bibitem[Mynick {\em et~al.\/}(1982)Mynick, Chu \& Boozer]{Mynick:1982fg}
{\sc \au{Mynick, H.~E.}, \au{Chu, T.} \& \au{Boozer, A.~H.}} \yr{1982}
  \at{{Class of Model Stellarator Fields with Enhanced Confinement}}.
  \jt{Physical Review Letters}  \bvol{48}~(5),  \pg{322--326}.

\bibitem[Mynick {\em et~al.\/}(2010)Mynick, Pomphrey \& Xanthopoulos]{Mynick_1}
{\sc \au{Mynick, H.~E.}, \au{Pomphrey, N.} \& \au{Xanthopoulos, P.}} \yr{2010}
  \at{{Optimizing Stellarators for Turbulent Transport}}.  \jt{Physical Review
  Letters}  \bvol{105}~(9),  \pg{095004}.

\bibitem[Spong {\em et~al.\/}(1998)Spong, Hirshman, Whitson, Batchelor,
  Carreras, Lynch \& Rome]{1998PhPl....5.1752S}
{\sc \au{Spong, D.~A.}, \au{Hirshman, S.~P.}, \au{Whitson, J.~C.},
  \au{Batchelor, D.~B.}, \au{Carreras, B.~A.}, \au{Lynch, V.~E.} \& \au{Rome,
  J.~A.}} \yr{1998}  \at{{J* optimization of small aspect ratio
  stellarator/tokamak hybrid devices}}.  \jt{Physics of Plasmas}  \bvol{5}~(5),
   \pg{1752--1758}.

\bibitem[Strand \& Houlberg(2001)]{Strand:2001cd}
{\sc \au{Strand, P.~I.} \& \au{Houlberg, W.~A.}} \yr{2001}  \at{{Magnetic flux
  evolution in highly shaped plasmas}}.  \jt{Physics of Plasmas}  \bvol{8}~(6),
   \pg{2782}.

\end{thebibliography}

\end{document}